\documentclass{article}
\usepackage{amsmath}
\usepackage{amssymb}\usepackage{amsfonts}\usepackage{float}
\def\nn{\nonumber}
\def\+{\pm}
\def\half{{\textstyle{1\over2}}}
\let\a=\alpha \let\b=\beta  \let\d=\delta 
    
\let\l=\lambda \let\m=\mu \let\n=\nu  \let\p=\partial 
    \let\c=\chi 
         
  \let\S=\Sigma  
\let\C=\Chi

\def\nn{\nonumber} \def\bd{\begin{document}} \def\ed{\end{document}}
\def\ds{\documentstyle} \let\fr=\frac \let\bl=\bigl \let\br=\bigr
\let\Br=\Bigr \let\Bl=\Bigl 
\let\bm=\bibitem
\let\na=\nabla
\let\pa=\partial \let\ov=\overline 
\def\p{\partial}
\def\m{\mu}
\def\n{\nu}
\def\S{\Sigma}
\def\a{\alpha}
\def\b{\beta}
\def\c{\gamma}
\def\d{\delta}
\def\l{\lambda}
\def\br{{\bf r}}
\def\bt{{\bf t}}
\def\be{{\bf e}}
\def\bn{{\bf n}}
\def\bb{{\bf b}}
\def\pa{\partial}
\def\bx{{\bf x}}
\def\ba{{\bf a}}
\def\bv{{\bf v}}
\def\bF{{\bf F}}
\def\bG{{\bf G}}
\def\bb{{\bf b}}
\def\by{{\bf y}}
\def\bt{{\bf t}}
\def\bu{{\bf u}}
\def\bp{{\bf p}}
\def\bL{{\bf L}}
\def\bB{{\bf B}}
\def\bE{{\bf E}}
\def\bH{{\bf H}}
\def\bD{{\bf D}}
\def\bS{{\bf S}}
\def\bk{{\bf k}}
\def\bc{{\bf c}}

\def\ft#1#2{{\textstyle{{\scriptstyle #1}\over {\scriptstyle #2}}}}
\def\fft#1#2{{#1 \over #2}}
\def\del{\partial}
\def\vp{\varphi}
\def\st#1{{\scriptstyle #1}}
\def\sst#1{{\scriptscriptstyle #1}}

\def\oneone{\rlap 1\mkern4mu{\rm l}}
\def\td{\tilde}
\def\wtd{\widetilde}
\def\ie{{\it i.e.\ }}
\def\dalemb#1#2{{\vbox{\hrule height .#2pt
        \hbox{\vrule width.#2pt height#1pt \kern#1pt

                \vrule width.#2pt}
        \hrule height.#2pt}}}
\def\square{\mathord{\dalemb{6.8}{7}\hbox{\hskip1pt}}}

\def\cramp{\medmuskip = 2mu plus 1mu minus 2mu}
\def\cramper{\medmuskip = 2mu plus 1mu minus 2mu}
\def\crampest{\medmuskip = 1mu plus 1mu minus 1mu}
\def\uncramp{\medmuskip = 4mu plus 2mu minus 4mu}

\newcommand{\ho}[1]{$\, ^{#1}$}
\newcommand{\hoch}[1]{$\, ^{#1}$}
\newcommand{\bea}{\begin{eqnarray}} 
\newcommand{\eea}{\end{eqnarray}} 
\newcommand{\ra}{\rightarrow}
\newcommand{\lra}{\longrightarrow}
\newcommand{\Lra}{\Leftrightarrow}
\newcommand{\ap}{\alpha^\prime}
\newcommand{\tr}{{\rm tr} }
\newcommand{\Tr}{{\rm Tr} } 
\def\0{{\sst{(0)}}}
\def\1{{\sst{(1)}}}

\def\2{{\sst{(2)}}}
\def\3{{\sst{(3)}}}
\def\4{{\sst{(4)}}}
\def\5{{\sst{(5)}}}
\def\6{{\sst{(6)}}}
\def\7{{\sst{(7)}}}
\def\8{{\sst{(8)}}}
\def\n{{\sst{(n)}}}
\def\cA{{{\cal A}}}
\def\cF{{{\cal F}}}
\def\tV{\widetilde V}
\def\tW{\widetilde W}
\def\tH{\widetilde H}
\def\tE{\widetilde E}
\def\tF{\widetilde F}
\def\tA{\widetilde A}
\def\im{{{\rm i}}}
\def\jm{{{\rm j}}}
\def\km{{{\rm k}}}

\def\tY{{{\wtd Y}}}
\def\ep{{\epsilon}}
\def\vep{{\varepsilon}}
\def\R{\rlap{\rm I}\mkern3mu{\rm R}}
\def\R{{{\mathbb R}}}
\def\C{{{\mathbb C}}}
\def\H{{{\mathbb H}}}
\def\CP{{{\mathbb C}{\mathbb P}}}
\def\RP{{{\mathbb R}{\mathbb P}}}

\def\bx{{\bf x}}
\def\wtd{\widetilde}
\def\ft#1#2{{\textstyle{{\scriptstyle #1}\over {\scriptstyle #2}}}}
\def\fft#1#2{{#1 \over #2}}
\def\by{{\bf y}}
\def\bx{{\bf x}}
\def\del{\partial}
\def\nn{\nonumber}
\def\sst#1{{\scriptscriptstyle #1}}
\def\0{{\sst{(0)}}}
\def\1{{\sst{(1)}}}
\def\2{{\sst{(2)}}}
\def\3{{\sst{(3)}}}
\def\4{{\sst{(4)}}}
\def\5{{\sst{(5)}}}
\def\6{{\sst{(6)}}}
\def\7{{\sst{(7)}}}
\def\8{{\sst{(8)}}}

\def\half{{1\over 2}} 
\def\ben{\begin{equation}}
\def\een{\end{equation}}

\begin{document}

\title { What is the Shape of a Black Hole?} 

\author{G.W. Gibbons \\
University of Cambridge, DAMTP, CMS,\\ Wilberforce Road, Cambridge CB3 0WA, UK\\}

\maketitle
\section{Introduction}
The basic problem I wish to tackle in this lecture \footnote{Written version of an invited lecture   given at the  
the International  Fall  Workshop  on Geometry  and  Physics  
held   In ICMAT, MADRID (Spain), from August  31 to  Sepstember 3 2011
and to appear in the Proceedings. }   is:

\medskip 
{{How  do  we  describe  and  
characterize  the   geometry
 of  horizons,  and  what  are  their  
general  geometrical  properties?} }  

\medskip  {
By ``horizon'' I shall mean  ``apparent horizon''
 or `` outermost closed  marginally trapped'' 
$D-2$ dimensional hypersurface  $S$  in a $D$ dimensional 
 asymptotically flat
or asymptotically  Anti-De-Sitter spacetime whose energy momentum tensor
satisfies the Dominant Energy Condition or (if $\Lambda <0$), some suitable
generalization. 
 \subsection{ Closed Trapped and Marginally Trapped  Surfaces}
A spacelike codimension-two surface has a timelike normal 2-plane
which contains two future directed null directions.
The local area elements of a  
{\it closed trapped  2-surface}  decreases
in both the inward
{\sl and} the outward null directions directions if pushed to the future along its
two lightlike normals
An {\it apparent horizon}  is a marginally outer trapped surface:
the local area  elements just fail to increase in the
 outward null direction.
As an {\sl example} 
suppose that $S$ lies in a Cauchy  surface
$\Sigma$ , with vanishing second fundamental form $K_{ij}$.  
Then $S$ is the outermost minimal surface in an asymptotically
flat Riemannian manifold with non-negative scalar  curvature
$R=R^i_i \ge 0$. 
The standard example is  the surface $\rho=\frac{E}{2} >0$ 
in  the Cauchy surface for the Schwarzshild Black Hole
\ben
ds ^ 2= \bigl(1 + \frac{E}{2\rho}  \bigr )
 ^4 \bigl ( d \rho ^2 + \rho ^2  ( d\theta ^2 + \sin ^2 \theta d \phi ^2  )
\bigr )  
\een 
in isotropic coordinates $\rho= \half (r-E) + \half \sqrt{r(r-2E)} $ \footnote
{ $g_{00}=-\frac{(1-\frac{E}{2\rho})^2}{(1+ \frac{E}{2\rho}) ^2 }$. $E$ is the ADM mass or total energy of the black hole. 
The metric has vanishing scalar curvature } . 

The many  exact solutions
of supergravity and Kaluza-Klein type theories in diverse dimensions
now available satisfy this condition
and provide  a copious 
supply  of examples for formulating and testing conjectures
In many cases, especially if $D\le 5$ and 
the cosmological constant $\Lambda =0$ ,   powerful
solution generating techniques, now often thought of  as T and S
dualities,   are available because  
of the high degree of symmetry of the space of  fields,  
which is not infrequently a {\it symmetric space} 
Following a paper by myself \cite{me} 
 Mirjam Cvetic, Chris Pope
 and I , with some initial help from a Summer Student, Thiti Sirithanakorn 
we \cite{we}  have been availing ourselves of these opportunities 

Note that supergravity solutions are useful in this way even if 
one is not interested in supergravity {\sl per se}.
\subsection{  What is meant by Shape ?} 
By ``shape '' I shall mean  {\sl intrinsic} geometrical properties,
such as the area $A$ which determined by the induced 
Riemannian metric $g$. The {\sl extrinsic}
geometry is determined by its being marginally trapped.
I shall also  be interested  in how the shape is   related
to dynamical  quantities such as the total energy $E$ ( i.e. ADM or
Abbott-Deser mass), total angular momentum $J$, or total angular  
momenta  $J_i$,
$i=1,2,\dots,[\frac{D-2}{2}]$ or electric charge $Q$ or charges $Q_a$.
 \subsection{ The Penrose Inequality} 
The best  known and best investigated example
of what I have in mind is variously called the {Penrose,
 Cosmic Censorship,  or Isoperimetric Inequality for Black Holes}
 In four dimensions\footnote{we set Newton's $G=1$} 
\ben \boxed{
A \le 16 \pi E^2}  \label{cosmic}
\een
Physically, since the  Bekenstein-Hawking    black hole entropy
\ben
\boxed{
S = \frac{1}{4} A }\,, \label{entropy}
\een
this is the statement that for fixed energy, the Schwarzschild, or the
Kottler solution has the largest possible entropy. However 
(\ref{entropy}) is only rigourously established for {\sl stationary},
i.e. time independent black holes.
All known examples are consistent
with this, and in the four-dimensional {\sl time symmetric} , or so-called 
{\sl Riemannian}  case
there are  rigourous proofs for asymptotically flat 
initial data due to Huisken and Ilmanen
and by Bray.
In higher dimensions we have work by Barrabes and Frolov and Gibbons
and Holzegel and by Bray. These provide  partial results showing that     
\ben \boxed{
\frac{A}{{\cal A}_{D-2} }
 \le \Bigl ( \frac{16 \pi E} {(D-2) {\cal A}_{D-2} } \Bigr) 
 ^{\frac{D-2}{D-3} }}  \een
where  $ {\cal A}_{n} $ is the volume  of a unit $n$-sphere.  
Bray's results on the Riemannian case   are valid up to $D=8$,
this seems to be  related to the  the failure of regularity of 
minimal surfaces
which also been encountered in Brane theory \cite{Kei} 
For {{\it axisymmetric}  initial data sets} , Dain and Reiris
\cite{Dain}  
have recently established that if $J$ is total angular momentum,
then  for the outermots mininal surfces in maximal initial data 
\ben \boxed{
A \ge 8 \pi |J|,} 
\een
with equality for extreme Kerr. An extension to stable marhinally
outer trapped surfces has been obtained by Jarmillo, Reiris and Dain
\cite{Jaramillo}. 
Combining  with the  Penrose conjecture one might  conjecture  that for 
{{\it axisymmetric}  initial data sets}
\ben \boxed{
E \ge \sqrt{\frac{|J|}{2}} \,.} 
\een
\subsection{ Two other measures of shape} 
If $D=4$, in addition to the area the metric $g$ on $S$  
gives rise to two other
important measures of the shape
\begin{itemize} \item The length $l(S,g)$ of the shortest
non-trivial closed geodesic
\item Birkhoff's invariant $\beta(S,g)$ 
 \end{itemize} 
There is an obvious  generalisation of $l(S,g)$ to higher dimensions 
the generalisation of  $\beta(S,g)$ is much less so
and not unique. 
\subsection{ Antipodal Isometries}
The horizons of all known isolated black holes   
in all dimensions admit an {\sl antipodal isometry}, that is 
fixed point free involution $I$ preserving the metric. E.g. $ I: (\theta, \phi) 
\rightarrow (\pi-\theta, \phi +\pi) $ for Kerr-Newman. 
If $D=4$, and this is assumed, then Pu showed,  
by passing to $S/I \equiv RP^2$,      
\ben
\boxed{ l(S,g) \le \sqrt{\pi A} } 
\een
Following  Berger and Gromov,  Geometers
refer to such inequalities as  {\it Systolic}. 
Assuming the Cosmic-Censorship Inequality (\ref{cosmic}) we obtain
\ben
\boxed{l(S,g) \le 4 \pi E  } \label{syt}
\een
which smells like the {\it Thorne's  Hoop Conjecture} ( see later)
\subsection{ A new conjecture} 
{I conjecture that if $D=4$, then}  
\ben
\boxed{l(S,g) \le 4 \pi E  } 
\een
{always holds
regardless of whether $\{S,g\}$ admits an antipodal isometry}  

 We have  checked it on all examples known to us. 
\subsection{ Higher Dimensions}In higher dimensions 
there is no general theorem analogous to Pu's theorem.
However, since in all cases examined 
 $S$ admits an antipodal isometry $I$  and
we have been able to identify and bound the length of 
closed geodesics invariant under $I$ we 
have obtained an upper  bound for  length
of any closed geodesic on the two-fold cover and hence for $l(g)$. 
It turns out that  that in all    even dimensional cases  examined that  
\ben \boxed{
l(S,g) \le 2 \pi  \Bigl (  \frac{A}{{\cal A}_{D-2}} \Bigr ) 
^ {\frac{1}{D-2}} } \een \ben \boxed{
l(S,g) \le \Bigl(\frac{16 \pi ^{D-2} E }{ (D-2) {\cal A}_{D-2} }   \Bigr )
^{\frac{1}{D-3}}  }
\een  \
For  all odd  $D>3$ we have found 
\ben \boxed{
l(S,g) \le \Bigl(\frac{16 \pi ^{D-2} E }{ (D-2) {\cal A}_{D-2} }   \Bigr )
^{\frac{1}{D-3}}  }
\een
but our results on 

\ben \boxed{
l(S,g) \le 2 \pi  \Bigl (  \frac{A}{{\cal A}_{D-2}} \Bigr ) 
^ {\frac{1}{D-2}} } \een 
are inconclusive, and we strongly  suspect that it fails.
 \section{Birkhoff's Invariant
and Thorne's Hoop Conjecture} 
To connect with Thorne's  Hoop conjecture
we turn to the Birkhoff invariant
 $\beta(S,g)$ in $D=4$ spacetime dimensions.
 We consider a  
{ ``foliation'', ``sweep out'' or
``slicing `` }
of $S$ by $S^1$ leaves $f={\rm constant}=c $ whose  
length or circumference  is $l(S,g,f, c)$  with two point-like leaves.
E.G. $f=\cos \theta\,, l= r_+ \sin \theta $ for Kerr-Newman. Let
$\beta(S,g,f)$ be the {\sl maximum}  circumference for that
choice of slicing
\ben
\beta(S,g,f) = \max _c \quad l(S,g,f,c)
\een
Now {\sl minimize} over all choices of slices (e.g.
``over all ways  of passing a hoop or an elastic band
 over the horizon'' ) and define
\ben
\beta(S,g) = \min _f \quad \beta(S,g,f) = \min_f \, \max _c l(S,g,f,c)
\een 
In 1917 \cite{Birkhoff} Birkhoff used the ``Mountain Pass  Method'' method to
prove that every metric on $S^2$ admits at least one non-trivial closed
geodesic whose length is no greater than $\beta(S,g)$ , thus
\ben
\boxed{l(S,g) \le \beta (S,g) \,.}
\een  
\subsection{ Another conjecture}
{In the spirit, if not perhaps not precisely 
the letter, of  Thorne's Hoop
conjecture I have recently conjectured that} \cite{me}
\ben
 \boxed{
\beta(S,g) \le 4 \pi E \,. } 
\een 
This implies the previous inequality
\ben
 \boxed{
l(S,g) \le 4 \pi E \,. } 
\een 
Obtaining an upper bound for $\beta(S,g)$ merely  entails
estimating the maximum circumference of a conveniently chosen foliation.
In all cases we have examined (with or without a negative $\Lambda$ term)
the conjecture has been verified.  These now extend to
rotating asymptotically flat   black holes with four distinct charges
and  rotating ADS black holes with two charges set equal.
Further evidence comes from collapsing shells. 
\section{ Collapsing Shells and Convex Bodies}   
This is  a class of examples \cite{Shell}   in which a shell of null matter
collapses at the speed of light in which the apparent horizon
$S$ may be thought of as a convex body isometrically embedded
in  Euclidean space ${ E} ^3$. In this case one has 
\ben
8 \pi M_{\rm ADM} \ge         \int _S H d A \,, \label{shell}
\een
where $H= \half ({1 \over R_1} + {1 \over R_2 } ) $ is the mean curvature
and $R_1$ and $R_2$ the principal radii of curvature
of $S$  and $dA$ is the area element on $S$.
The right hand side is called the total mean curvature and 
it was  shown by \'Alvarez Paiva  \cite{Paiva}   
 that  in this case that
\ben
\beta(g) \le  \half \int _S H d A \,. \label{Paiva}  
\een
Combining \'Alvarez Paiva's (\ref{Paiva})  with (\ref{shell}) establishes
the conjecture  in this case.

Surprisingly, perhaps my conjecture holds up
even if  in grossly non-asymptotically flat situations
with a magnetic field. However, as we see, this also leads to an
apparent difficulty with Thorne's formulation. 

In fact the proof of \'Alvarvarez-Paiva is close to the ideas of Tod.
If ${\bf n}$ is a unit vector
we define the height function on $S \subset { E} ^3$ by
\ben
h= {\bf n}.{\bf x} \,, \qquad {\bf x \in S}\,.  
\een
Let $S_{\bf n}$ be the orthogonal
projection of the body $S$ onto a plane 
 with unit normal ${\bf n}$ and let $C( {\bf n}) =l( \partial
 S_{\bf n} ) $ be the 
perimeter of  $S_{\bf n}$. 
Then
\ben
\beta (g) \le \beta(g,h) \le C({\bf n}) \,, \label{height}  
\een
where, as $\beta(g)$ is the Birkhoff in invariant
for the induced metric on the convex surface and $\beta(g,h)$ 
is its  estimate using the foliation by the height function.  Now 
\ben
\int _S H d A = { 1 \over 2 \pi} \int _{S^2} C({\bf n}) d \omega \,,
\label{cross} 
\een 
where $d \omega$ is the standard volume element on the round two-sphere 
$S^2$ of unit radius. Thus averaging (\ref{height}) over $S^2$ and using 
(\ref{cross}) gives (\ref{Paiva}).
\subsection {Asymptotically-Melvin black holes}
were first constructed using a Harrison transformation
in Einstein-Maxwell theory,
by   Ernst  and in an explicit form by 
Ernst and Wild .
The metric is 
\ben
ds_4^2 = F^2 \Bigl\{ -\bigl (1-{2E \over r}\bigr ) dt^2 +  
{ dr^2   \over 1-{2 E \over r} }    + r^2 d\theta^2  \Bigr\}
+ {r^2 \sin^2\theta \over F^2 }  d\phi^2 \,,  
\label{Ernst}
\een
with 
\ben
F = 1 + {B^2 \over 4} r^2 \sin^2\theta\,,    
\een
where $B$ is the applied magnetic field.
If $E=0$ we get the Melvin solution, whilst if instead $B=0$ we get the
Schwarzschild solution. The energy with respect
to the Melvin background is  $E$
and the horizon, which is located at $r=2E$. 
{\
If $\gamma =E\, |B|$, the horizon  metric is
\ben
ds^2 = 4 E^2 \Bigl \{
 (1+ \gamma^2 \sin^2\theta)^2  d\theta^2 + {\sin^2\theta \over 
(1+ \gamma^2 \sin^2\theta )^2 }  d\phi^2  \Bigr \} \,,   
\een
Remarkably, these are the same as in the absence of the magnetic field.
\ben
\beta(g) \le \max  _\theta \, \frac{4\pi\, E\sin \theta}{1+ \gamma^2
  \sin^2\theta} 
\een
If $\gamma \le 1$, the circumference  $C(\theta)$ has a single maximum
at the equator $\theta =\frac{\pi}{2} $, the maximum value being
$\frac{4  \pi E }{1+\gamma^2}  \le 4 \pi E$. If $\gamma \ge 1$, the horizon is
dumb-bell shaped
and has two maxima with  $\gamma \sin \theta =1$, the maximum value
being $\frac{2E\pi}{ \gamma} < 4 \pi E$. Thus the conjecture is always
satisfied. 

The solutions for a black hole immersed in a
magnetic field in Einstein-Maxwell-Dilaton theory
have been given by Yazadjiev. The conjecture 
continues to hold. 
\subsection { Problems with Thorne's  Hoop conjecture?}
This was that \cite{Thorne}
\medskip

{\it
 Horizons form when and only when a mass $E$ gets
compacted onto a region whose circumference in EVERY direction is
$C\le 4 \pi E$. }

\medskip 
The capitalization  \lq \lq EVERY \rq \rq was 
intended to emphasis the fact that while the collapse of 
oblate shaped bodies,   the circumferences are all roughly equal,
in the  prolate case, a the collapse of a long almost cylindrically 
shaped body whose girth was never the less small would not necessarily
produce a horizon.  
However the  polar circumference is 
\ben
C_p= 4 E \int _0^\pi (1+ \gamma ^2 \sin ^2 \theta ) \,d \theta 
= 4 \pi E(1+ \half \gamma  ^2) \ge 4 \pi E \,. 
\een

{\sl This would seem to contradict Thorne's  ``in all directions'' 
formulation of the Hoop
Conjecture}.

\subsection{Isometric Embeddings}
One way of visualizing two dimensional surfaces
is to globally embed them isometrically into three dimensional Euclidean space
$E^3$. If the Gauss  curvature is everywhere positive, 
then by a theorem  of Weyl and Pogorelov this is always possible
and the  embedding is unique. 
Thus  no  ambiguity results from  such  
{``inflexible''  embeddings}.
Contrary to a statement by Ernst and Wild, the horizon
of the Ernst Wild black hole  can be globally isometrically embedded into 
Euclidean space even though its Gauss curvature can become
negative near the waist of the ``dumb bell''. 
However
despite being prolate, the Gaussian curvature $K$ of the 
horizon of  Kerr-Newman black hole is 
\ben
K= { ( {r_+} ^2 + a^2 )  
( r_+ ^2 -3 a ^2 \cos ^2 \theta )  
\over ( { r_+} ^2 + a ^2 \cos ^2 \theta )  ^3 } 
\,,  
\een 
$K$  can become negative at the poles $\theta=0, \pi$
and  this precludes a global  isometric embedding into $E^3$ 
as discovered by Smarr \cite{Smarr}.

Frolov has pointed out that one may globally embed into four dimensional
 Euclidean space $E^4$ , but this is probably not unique.
However a theorem of Pogorelov guarantees a {\sl unique} isometric
embedding into three dimensional hyperbolic space $H^3$.
This may be easily  achieved  using the upper half space model
for $H^3$ \cite{Global}.
\section{ Hyper-Hoops}
The analogue of a ``hoop'' is a `` hyper-hoop'', 
a bag or surface of one less dimension than the horizon which can be 
``dragged'' over it. Thus we have a sweep out or foliation by a one parameter
family of $D-3$ dimensional surfaces, each of which has an area.
In any given foliation $f$ we set  
\ben
\beta(S,g,f) = \max_c  A_{D-3} (f^{-1}(c)) 
\een
and define  
\ben
\beta(S,g) = \min _f \beta (S,g,f)
\een

$\bullet$Such sweep-outs have been used by mathematicians  to construct minimal 
surfaces via the mountain pass method.
 \
If we consider topologically spherical horizons $S \equiv S^{D-2}$,
then obvious choices for ``hyper-hoops''  are $S^p \times S^q$, $p+q= D-3$. 
E.G. on a round sphere
\ben
ds ^2 = d \theta ^2 + \sin ^2 \theta d\Omega^2_p + \cos ^2 \theta d \Omega_q ^2 
\een

If $ q=0$, we let $0\le \theta \le \pi$. If $pq \ne 0$  we let
 $0\le \theta \le \frac{\pi}{2}$.

$\bullet$ For Myers-Perry-AdS  black holes with two unequal angular momenta
$J_1 \ne J_2$ 
in $D=5$, we can choose $p=q=1$ and use the toroidal hyper-hoops
swept out by the $U(1)\times U(1)$ rotational  sub-group.
For $ {\rm Tangherlini}_5$ , these hyper-hoops are Clifford Tori.
For Clifford sweep outs we find
\ben \boxed{
\beta(S,g) \le \frac{16 \pi }{3} E \,.}  
\een   
This agrees with some earlier numerical work 
using the time symmetric initial value problem  \cite{Ida,Yamada}
Ida and Nakao also pointed out that in $D\ge 5$  
some circumferences may become extremely long, and so Thorne's
in all directions conjecture fails \cite{Ida} . 
  \subsection{ Sweep outs by $S^1 \times S^{D-4} $}
$\bullet$ If we consider the odd dimensional Kerr-AdS solutions
with all angular momenta equal $J_1=J_2=\,\dots =J_{ [\frac{D-1}{2}]}   $   
then the high  $SU([{D-1}{2}  ]  )$ symmetry  allows us to foliate  
by hyper-hoops with   topology $S^1 \times S^{D-4} $.
We find the obvious generalization of
the conjecture for the Birkhoff invariant holds.
 We can also consider just one non-vanishing angular momentum.
This also allows an external magnetic field. If $D= 2N+1$ is odd, we find 
\ben \boxed{
\beta(g)  \le  \fft{32\pi}{(2N-1)}\, (N-1)^{\ft12(N+1)}\, N^{-\ft12 N}\,            E \,, }
\een
\subsection{Diameters} One might have thought, particularly  because
the of the existence  of antipodal maps that the 
diameter $d(S,g) = \sup_{x,y \in S} d(x,y) $ 
might be useful. However the relation between
$d(S,g)$ and $l(S,g)$ is rather obscure, even for metrics on $S^2$
as is clear from  recent paper \cite{Croke} 
whose abstract reads  {\it  We construct
counterexample to a conjectured inequality $L\le    2D$ relating
the diameter $D$ and the least length $L$ of a nontrivial 
closed geodesic, for a Riemannian metric on the 2-sphere. 
The construction relies on Guillemin's
theorem concerning the existence of Zoll surfaces integrating an arbitrary
infinitesimal odd deformation of the round metric. Thus the round metric is
not optimal for the ratio $L/D$.}
According to these authors  one knows for sure that $L \le 4D$ 
\section{Possible Future Directions } 
\subsection{ Can one hear the shape of 
a black hole ? } 
The spectrum of the Laplacian on $\{S,g\}$  has been studied
recently in $D=4$ spacetime dimensions  in the spirit
of Kac's well known question
\cite{Engman1,Engman2} }.As far as one can
see,  it does not tell one much about the external
geometry since a two family of metrics is common to
Kerr-Newman and its SUGRA generalization to two charges.  
{Note this is} {\bf not} { the same as determining the 
black hole parameters  from the  spectrum of quasi-normal modes
which seems to be quite feasible}

The Kerr-Newman metrics have been generalized to include up
to four different charges associated with four different abelian vector
fields \cite{Youm}  In the subclass for which
only two charges are non-vanishing, from  the results,
of \cite{Chong} the horizon metric is  
\ben
ds ^2 = W d \theta ^2 + { (r_{+1}r_{+2} + a^2 )^2 \over W } 
\sin ^2 \theta  \,d \phi ^2 \,, 
\een

with
\ben
W= r_{+1}r_{+2} + a^2 \cos^2 \theta \,.\
\een
and
\ben
 r_{+1}= r_+ + 2  m \sinh ^2  \delta  _1 \,, 
\qquad  r_{+2}= r_+ + 2 m \sinh ^2  \delta _2
\een
with $r_+$ the larger root of $r^2 -2mr + a^2 =0$
and $\delta _1 $ and $\delta_2$ two parameters specifying the two charges.
If $\delta_1=\delta _2$ we obtain the Kerr-Newman case.

The horizon  metric  depends only on two parameters
$r_{+1}r_{+2}$ and $a$ which in this case Engman et al. showed
may  be determined
from the spectrum.  The metric is isometric to
that of  the neutral Kerr spacetime.
However, the spacetime  metric depends on the mass, angular momentum and two charges. 
Of course the interpretation of the parameters occurring in the horizon metric
is different,
but the geometry is the same.

\subsection{What is the volume of
a black hole ?}.
In another recent paper \cite{Cvetic}  
 we defined the {Thermodynamic Volume}
of a black hole by calculating the {enthalpy $E$ 
 of a black hole}. Note that the enthalpy and energy coincide of the 
pressure or volume are kept constant and so  we have indulged in a
slight abuse of notation and used the same symbol $E$ for both.
In fact ss boundary integrals at infinity  they are given by 
the same expressio 
The first law with the cosmological constant $\Lambda$ or  pressure
$P= -\frac{D-2}{16 \pi} \Lambda$ varied reads
\ben
dE = TdS + \Omega _i d J_i + \Phi_a dQ_a + \Theta d \Lambda
\label{law}\een
where the thermodynamic volume $V= - \frac{16 \pi \Theta }{D-2}$
and $\Theta$ is defined by the first law (\ref{law})  . 
Surprisingly we found that $V$ satisfies a 
{\it Reverse Isoperimetric Inequality}
\ben
\Bigl( \frac{(D-1)V }{{\cal A}_{D-2}} \Bigr )  ^{\frac{1}{D-1}}
\Bigl( \frac{{\cal A}_{D-2} } {A}\Bigr) ^ {\frac{1}{D-2} }  \ge 1.
 \een 
Curiously one may take the limit $\Lambda \rightarrow 0$
to get an expression for $V$.

\section{Conclusions}

\begin{itemize}
\item If $D=4$ , there is good evidence and some 
proofs in special cases that  
\ben
 \boxed{
\beta(S,g) \le 4 \pi E \,. } 
\een 
which implies \ben
 \boxed{
l(S,g) \le 4 \pi E \,. } 
\een 
which follows if
\ben
\boxed{
l(S,g) \le \sqrt { \pi A}   \qquad {\rm and}\qquad  
\sqrt { \pi A}  \le 4 \pi E  \,.      } 
\een

\item If $D$ is  even, for all examples tested   
\ben \boxed{
l(S,g) \le 2 \pi  \Bigl (  \frac{A}{{\cal A}_{D-2}} \Bigr ) 
^ {\frac{1}{D-2}} } 
\een 
\ben 
\boxed{
l(S,g) \le \Bigl(\frac{16 \pi ^{D-2} E }{ (D-2) {\cal A}_{D-2} }   \Bigr )
^{\frac{1}{D-3}}  }
\een
The second inequality holds in odd $D>3$ while the first
seems to fail

\item If $D=5$  sweep outs by 2-tori satisfy  
\ben \boxed{
\beta(S,g) \le \frac{16 \pi }{3} E \,.}  
\een  
$\bullet$ If $D=2N+1$ sweep outs by $S^1 \times S^{D-4}$ satisfy 

\ben \boxed{
\beta(g)  \le  \fft{32\pi}{(2N-1)}\, (N-1)^{\ft12(N+1)}\, N^{-\ft12 N}\,            E \,, }
\een

\item These inequalities continue to hold in the
presence of external Melvin type 
magnetic fields in $D=4$ and higher dimensions  

\item  However these examples, and higher dimensional
rotating black holes seem to invalidate the   ``in all directions''
part of Thorne's conjecture.

\item If $D=4$, Rapidly rotating horizons may not be globally
isometrically  embedded in Euclidean space $E^3$ , but
they can be globally
isometrically  embedded in Hyperbolic  space $H^3$

\end{itemize}

\end{document}